\documentclass[a4paper,10pt]{article}
\usepackage{graphicx}
\usepackage[left=2cm,right=2cm,
top=2cm,bottom=2cm,bindingoffset=0cm]{geometry}
\usepackage{amssymb}
\usepackage{amsmath}
\usepackage[classfont=bold,langfont=roman,funcfont=roman]{complexity}
\usepackage{cite}
\usepackage{xcolor}
\usepackage{multicol}
\usepackage{mathtools}
\def\be{\begin{eqnarray}}
\def\ee{\end{eqnarray}}

\title{Scattering solution of Schr\"odinger equation with  $\delta$-potential in deformed space with minimal length}
\author{M. I. Samar and V. M. Tkachuk\\ Professor Ivan Vakarchuk Department for Theoretical Physics, \\ Ivan Franko National University of Lviv,\\ 12 Drahomanov St, Lviv,
	UA-79005, Ukraine}
\begin{document}

\maketitle

\begin{abstract}
We consider the  Dirac $\delta$-function potential problem in the general case of deformed Heisenberg algebra leading to the minimal length.  Exact bound and scattering solutions of the problem in quasiposition representation are presented. We obtain that for some resonance energy  the incident wave is completely reflected. We conclude that this effect is very sensitive to the choice of the deformation function. 
\end{abstract}

\section{Introduction}
In the present paper we  consider  a  modified  one-dimensional  Heisenberg  algebra which is generated  by
position $\hat{X}$ and momentum $\hat{P}$ hermitian operators satisfying the following relation
\be \label{general_deformation}
[\hat{X},\hat{P}]=i\hbar f({\hat{P}}),
\ee
where $f$ is called the function of deformation. We assume that it is strictly positive ($f >0$), even function.

Algebra (\ref{general_deformation}) possesses the following  representation leaving position operator undeformed
\be\label{psevdo-position}
&&{\hat{X}}=\hat{x},\\ \nonumber
&&{\hat{P}=g({\hat{p}})},
\ee
where $\hat{x}$ and $\hat{p}$ satisfy non-deformed commutation relation 
\be
[\hat{x},\hat{p}]=i\hbar. 
\ee 
Depending on convenience,  operators $\hat{x}$ and $\hat{p}$  can be described using  coordinate
representation $\hat{x}=x$, $\hat{p}=-i\hbar\frac{d}{dx}$, or  momentum representation $\hat{x}=-i\hbar\frac{d}{dp}$, $\hat{p}=p$.

Function $g({p})$ from (\ref{psevdo-position}) is an odd function satisfying \be\frac{dg(p)}{dp}=f(P)\ee and defined on $[-b,b]$, with $b=g^{-1}(a)$. Here $a$ denotes the endpoint of momentum $P \in [-a,a]$.
It can be shown that if $b<\infty$ nonzero minimal length exists and if $b=\infty$ the minimal length is zero \cite{Maslowski}. 

One of the simplest deformed algebras is the one proposed by Kempf \cite{Kempf1994}
\be \label{deformation} [\hat{X},\hat{P}]=i\hbar(1+\beta
\hat{P}^2),\ee
leading to minimal length  $l_0=h\sqrt{\beta}$. 
By modifying usual canonical commutation relations Kempf et al. \cite{Kempf1994,KempfManganoMann,HinrichsenKempf,Kempf1997} showed that the effect of minimal length can be phenomenologically introduced in quantum mechanics, while on fundamental level minimal length as a finite lower bound to the possible resolution of length was suggested by string theory and quantum gravity\cite{GrossMende,Maggiore,Witten}.

One of the important problems concerning to minimal length  is to propose the effect or phenomenon on the basis of which the  minimal length hypothesis can be verified. Since in deformed space with minimal length coordinate representation does not exist, systems with singularity in potential energy are expected to have a  nontrivial sensitivity to the minimal length. From this point of view,  the  study of the effect of the minimal length on systems with singular potentials is of particular interest. 
 
 The influence of the minimum length  has been studied in the context of the following   problems with singularity in potential energy:   hydrogen atom \cite{Brau,Benczik,StetskoTkachuk,Stetsko2006,Stetsko2008, SamarTkachuk, Samar}, gravitational quantum well \cite{Brau2006, Nozari2010, Pedram2011}, a particle in delta potential and double delta potential\cite{Samar1, Ferkous}, one-dimensional Coulomb-like problem \cite{Samar1,Fityo,Samar2}, particle in the singular inverse square potential \cite{Bouaziz2007,Bouaziz2008, Bouaziz2017,Samar2020}.
 
 We organize the rest of this paper as follows. In Section 2, we discuss some peculiarities of quasiposition representation. In Section 3 the problem of definition of  $\delta$-potential in deformed space with minimal length is reviewed. Continuity equation with minimal length hypothesis is considered in Section 4. Bound solution 
of $\delta$ potential problem in quasiposition representation is presented in Section 5.
 While, in Section 6, scattering problem for $\delta$ potential is studied.
 Some concluding remarks are reported in the Section 7.

\section{Quasiposition representation}

An important feature of quantum theory with minimal length is that the eigenstates of
the position operator are no longer physical states, since for these states the uncertainty in position is zero, i.e. less then minimal. 
Technically, minimal uncertainty in position corresponding to (\ref{general_deformation}) means that the position operator is no longer essentially self-adjoint but has a one-parameter family of self-adjoint extensions \cite{KempfManganoMann, Samar2}.
As a result, the position
representation can no longer be used, that is, any wave function  can not be expanded in the basis of  eigenfunction of the position operator.  

However, the so-called quasiposition representation of deformed algebra (\ref{general_deformation}) can still be introduced  if we use  coordinate
representation  for small operators $\hat{x}=x$ and $\hat{p}=-i\hbar\frac{d}{dx}$. In this case representation of (\ref{general_deformation}) can be written as 
\be\label{quasiposition} \hat{X}&=&x,   \\
\nonumber \hat{P}&=&g\left(-i\hbar\frac{d}{dx}\right),
\ee
with $x\in(-\infty,\infty).$
An arbitrary wave function in  the quasiposition representation  can be presented as the following expansion  
\be\label{qcwf}
\psi(x)=\int_{-b}^bC(p)\varphi_p(x)dp
\ee
in the basis of eigenfunctions \be \varphi_{p}(x)=\frac{1}{\sqrt{2\pi\hbar}}
e^{i\frac{px}{\hbar}}, \ p \in [-b,b] \ee of self-adjoint operator $\hat{p}=-i\hbar\frac{d}{d
	x} $. 
Function $C(p)$ is the momentum representation wave function satisfying
\be\label{C}
C(p)=\int_{-\infty}^{\infty}\psi(x)\varphi^{*}_p(x)dx.
\ee
Substituting (\ref{C}) into  (\ref{qcwf}) we obtain  the condition of completeness
\be \label{property}
\psi(x)=\int_{-\infty}^{\infty}\psi(x')\tilde{\delta}(x-x')dx',
\ee
with
\be
\tilde{\delta}(x-x')=\int_{-b}^b\varphi^{*}_p(x')\varphi_p(x)dp=\frac{\sin\frac{\pi(x-x')}{2l_0}}{\pi(x-x')}.
\ee

We want to emphasize here that for the quasiposition wave function  given by (\ref{qcwf})  function $\tilde\delta$ possesses the property (\ref{property}) of Dirac delta function.

As an example let us obtain the  expression for the position operator eigenfunctions in quasiposition representation. 
In momentum representation these functions have the form 
\be \phi_{\lambda}(p)=\frac{1}{\sqrt{2b}}
e^{i\frac{\lambda p}{\hbar}},\ee
being the solutions of the following eigenequation
\be
i\hbar\frac{d\phi_{\lambda}(p)}{dp}=\lambda \phi_{\lambda}(p).
\ee
According to (\ref{property}) the position operator eigenfunctions in quasiposition representation writes
 \be \psi_{\lambda}(x)=\frac{\sqrt{2l_0}}{\pi}\frac{\sin\frac{\pi(x-\lambda)}{2l_0}}{(x-\lambda)}={\sqrt{2l_0}}\tilde\delta(x-\lambda).
 \ee

However, if we write the equation for the eigenvalues of the position operator $\hat{X}$ in the quasiposition representation, then we arrive to a  contradiction
\be\label{contr}
x \cdot \frac{\sqrt{2l_0}}{\pi}\frac{\sin\frac{\pi(x-\lambda)}{2l_0}}{(x-\lambda)}\neq\lambda \frac{\sqrt{2l_0}}{\pi}\frac{\sin\frac{\pi(x-\lambda)}{2l_0}}{(x-\lambda)}.
\ee
To resolve this problem,  we write the action of the position operator $\hat{X}=i\hbar\frac{d}{dp}$ in the momentum representation on an arbitrary wave function $ C (p) $ and make a transition to quasiposition representation
\be
\int_{-b}^b i\hbar \frac{dC(p)}{dp} \varphi_p(x)dp=x\psi(x)+i\hbar\left(C(b)\varphi_b(x)-C(-b)\varphi_{-b}(x)\right).
 \ee
This means that the coordinate operator in the quasiposition representation has the form
 \be
\hat{X}\psi(x)=x\psi(x)+i\hbar\left(C(b)\varphi_b(x)-C(-b)\varphi_{-b}(x)\right).
 \ee

By a direct substitution, it can be shown that the following definition of the coordinate operator solves the contradiction (\ref{contr}):
\be
\hat{X}\cdot \frac{\sqrt{2l_0}}{\pi}\frac{\sin\frac{\pi(x-\lambda)}{2l_0}}
{(x-\lambda)} = \lambda \frac{\sqrt{2l_0}}{\pi}\frac{\sin\frac{\pi(x-\lambda)}{2l_0}}{(x-\lambda)}.
\ee
Note that for physical states, i.e. states for which the mean value of kinetic energy is finite,
the condition $ C(\pm b) = 0 $ is required. In this case  
   \be
\hat{X}\psi(x)=x\psi(x),
 \ee
and the formula (\ref{quasiposition}) remains correct.
It is worth to note that for position operator eigenfunctions condition $ C(\pm b) = 0 $ is not satisfied.

\section{On defining the Dirac delta potential in deformed space with minimal length }

 Two different definitions of Dirac delta potential  in deformed space with minimal length are presented in literature. The first one corresponds to the assumption of the potential energy to be the same as in the undeformed case
\be \label{naive_delta} V(x)=-V_0\delta(x). \ee
Corresponding  Schr\"odinger equation can be written as follows
\be\label{incorrect_eqn}
\left(g^2\left(-i\hbar\frac{d}{dx}\right)-2mE\right)\phi(x)- 2mV_0\delta(x)\phi(x)=0.
\ee
Such equation was considered in \cite{Gusson} and earlier in \cite{Ferkous} in the linear approximation on deformation parameters in some special case of deformation function. 
However,  in our opinion, the problem is not correctly defined, since the first term in the left-hand side of (\ref{incorrect_eqn}) demanding $\phi(x)$ to be infinitely differentiable function is inconsistent with  the one containing  delta function. For more details see Appendix A. 
This fact shows that  the rash usage  of the approximation procedure on parameter of deformation can lead to incorrect results.  

In our previous paper \cite{Samar1}, we propose the definition of the delta potential assuming that the kernel of the potential energy operator does not change in the case of deformation. Thus, in momentum representation, the problem under consideration can be defined as
\be
\left(g^2\left(p\right)-2mE\right)\varphi(p)-\tilde V_0\int_{-b}^b\varphi(p) dp=0,
\ee
with \be\label{tV} \tilde V_0= \frac{mV_0}{\pi\hbar}.\ee
Making Fourier transformation we obtain
the Schr\"odinger equation in quasiposition representation
\be\label{correct_eqn}
\left(g^2{\left(-i\hbar\frac{d}{dx}\right)}-2mE\right)\phi(x)-2m V_0\tilde{\delta}(x)\hat{P_0}\psi(x)=0, 
\ee
which corresponds to the following definition for delta potential operator in quasiposition representation as   
\be\label{delta} \hat{V}(x)=-V_0\tilde{\delta}(x)\hat{P_0},\ee
with $\hat{P}_0$ denoting the operator the following projection operator $\hat{P}_0\psi(x)=\int_{-\infty}^{\infty}\tilde{\delta}(x)\psi(x)dx=\psi(0)$.
Here we emphasize that in the limit to undeformed case ($b\rightarrow\infty$) equation (\ref{correct_eqn}) leads to the usual Schr\"odinger equation for particle in delta potential.

It is also interesting to consider the following definition of $\delta$ potential
\be \hat{V}(x)=-V_0\tilde{\delta}(x),\ee
However, corresponding Schr\"odinger equation
\be\label{correct_eqn}
\left(g^2{\left(-i\hbar\frac{d}{dx}\right)}-2mE\right)\phi(x)-2m V_0\tilde{\delta}(x)\psi(x)=0, 
\ee
leads to integro-functional equations in momentum representation
\be
&&\left(g^2\left(p\right)-2mE\right)\varphi(p)-\tilde V_0\int_{-b}^{p+b}\varphi(p) dp=0, \ \ p\in[-b,0], \\
&&\left(g^2\left(p\right)-2mE\right)\varphi(p)-\tilde V_0\int_{p-b}^{b}\varphi(p) dp=0,\ \  p\in[0,b].
\ee
It is hard to achieve the exact solution of these equations. This problem deserves to be the subject of separate study in the future.  

\section{Continuity equation}
In paper \cite{Laba} the exact continuity equation in general case of deformed space was derived. For the system describing by the Hamiltonian 
\be
\hat H=\frac{g^2(\hat{p})}{2m}+V(x)
\ee
the following continuity equation was obtained
\be
\frac{\partial\rho}{\partial t}+\frac{\partial j}{\partial x}=0,
\ee
with \be
\rho=|\psi(x)|^2
\ee
and 
\be \label{j}
j=\frac{1}{2m}\sum_{n=1}^{\infty}a_n\sum_{k=1}^{n}(-1)^k[(\hat{p}^{k-1}\psi^*)\hat{p}^{2n-k}\psi-(\hat{p}^{k-1}\psi)\hat{p}^{2n-k}\psi^*],
\ee
with $a_n$ given by the series  $g^2(p)=\sum_{n=1}^\infty a_n p^{2n}$.

The introduced point interaction describing by Dirac delta function potential energy operator is not presented as operator of multiplication of some coordinate dependent function, but as a projection operator $ \hat{V}(x)=-V_0\tilde{\delta}(x)\hat{P}_0$. Therefore, the continuity equation has to be modified as

\be
\frac{\partial\rho}{\partial t}+\frac{\partial J}{\partial x}=0, 
\ee
where 
\be
J=j+j',
\ee
with  $j$ given in (\ref{j}) and
\be
j'=V_0\int_{-\infty}^x\tilde\delta(x')[\psi^*(x')\psi(0)+\psi^*(0)\psi(x)']dx.
\ee
It is important to note that in the limit of $x\rightarrow\infty$ the flow probability density $j'$ goes to zero. The flow probability density $j'$ can be considered as the one "inside" the point interaction and can be neglected "outside" the point interaction for $x>>l_0$, with $l_0$ denoting minimal length.
  
In case of plane wave $\psi(x)=Ce^\frac{ipx}{\hbar}$  in the limit of $x\rightarrow\infty$ flow probability density $J$ can be described by
 \be\label{flux}
 J=j=|C|^2\frac{\partial T(p,b)}{\partial p},
 \ee
with $T(p,b)=\frac{g^2(p)}{2m}$.

\section{Bound state solution of Schr\"odinger equation with delta potential in quasiposition representation }

In  the  Section 3 we arrive to the conclusion that  potential energy $V(x)=-V_0\delta(x)$ is inconsistent with deformed algebra with minimal length. 
Therefore we consider the following definition of the potential energy operator instead  $\hat{V}(x)=-V_0\tilde{\delta}(x)\hat{P_0}$. Let us find the bound state solutions of Schr\"odinger equation (\ref{correct_eqn}). 

In case of the bound states ($E<0$), we denote $q=\sqrt{-2mE}$ and rewrite Schr\"odinger equation (\ref{correct_eqn}) as the following
\be\label{correct_eqn_bound}
\left(g^2{\left(-i\hbar\frac{d}{dx}\right)}+q^2\right)\phi(x)-2m V_0\tilde{\delta}(x)\hat{P_0}\psi(x)=0. 
\ee
Solution of the equation (\ref{correct_eqn_bound}) we propose in the form
\be\label{correct_sol}
\psi(x)= \int_{-b}^{b}{\frac{Ae^{i px/\hbar}}{g^2(p)+q^2}dp}.
\ee
If we substitute $\psi(x)$ in  equation (\ref{correct_eqn}) we obtain the condition on the energy spectrum  
\be
1=\tilde V_0\int_{-b}^b\frac{1}{g^2\left(p\right)+ q^2}dp, 
\ee 
with $\tilde V_0$ given in (\ref{tV}).
In case of Kempf's deformation function
\be
f(P)=1+\beta P^2; \ \ g(p)=\frac{1}{\sqrt{\beta}}\tan(\sqrt{\beta} p), \ \  b=\frac{\pi}{2\sqrt{\beta}},
\ee 
 we obtain that the energy spectrum, like in undeformed case, consists of one energy level
\be E=-\frac{1+2\pi\tilde V_0\sqrt{\beta}-\sqrt{1+4\pi \tilde V_0\sqrt{\beta}}}{4m\beta}. \ee
For small $\beta$ energy spectrum can be approximated as
\be \label{Delta_energy}
E=-\frac{m V_0^2}{2\hbar^2}+\frac{m^2V_0^3}{\hbar^3}\sqrt{\beta}-\frac{5m^3V_0^4}{2\hbar^4}\beta+o(\beta^{3/2}).
\ee
Here we emphasise that this result up to the notations coincides with the one we  obtained earlier in \cite{Samar1}.

\begin{figure}[h]
	\centering
	\includegraphics[width=14 cm]{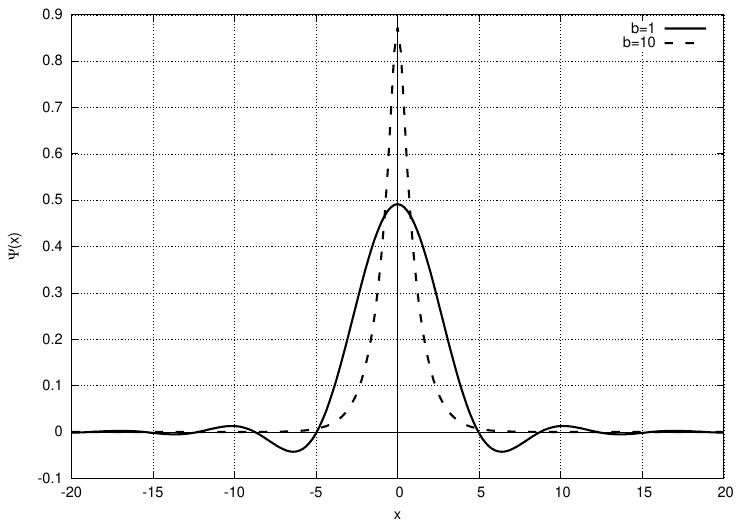}
	\vspace{-10 pt}
	\caption{\footnotesize{Eigenfunction of particle in delta potential for $b=1$ (solid) and $b=10$ (dashed) }}
	\label{fig1}
\end{figure}

The corresponding normalized eigenfunction is
\be\label{normalised phi}
\psi(x)= \sqrt{\frac{2}{\pi}}\frac{\beta(1+\sqrt{\beta}q)}{\sqrt{1+2\sqrt{\beta}q}}\int_{-b}^{b}{\frac{e^{i px/\hbar}q^{3/2}}{\tan^2\left(\sqrt{\beta}p\right)+\beta q^2}dp}.
\ee

From the  Fig.\ref{fig1}  we see that increasing parameter of deformation $b$ the eigenfunction  became more sharp at the origin reaching the features of undeformed eigenfunction.

\section{Scattering solution for $\delta$ potential with minimal length}
Let us consider the scattering solution ($E>0$) of Schr\"odinger equation (\ref{correct_eqn}) with  $\delta$-potential in deformed space with minimal length. If we denote $k=\sqrt{2mE}$  equation (\ref{correct_eqn}) writes
\be
\left(g^2\left(p\right)-k^2\right)\varphi(p)+\tilde{V}_0\int_{-b}^b\varphi(p) dp=0.
\ee
The solution of this equation can be written as
\be
\varphi(p)=\delta(p-k)+\frac{A}{g^2(p)-k^2},
\ee
with 
\be
A=-\frac{\tilde{V}_0}{1+\tilde{V}_0I(k)}.
\ee
Here we use notations
\be\label{I}
I(k)=\lim_{\varepsilon\rightarrow0}\int_{-b}^b\frac{1}{g^2(p)-k^2-i\varepsilon}dp= G(k) +\frac{i\pi}{kf(k)}
\ee
and
\be\label{pv} G(k)={\displaystyle {\mathcal {P}}} \int_{-b}^b\frac{1}{g^2(p)-k^2}dp.\ee
In the last formula ${\displaystyle {\mathcal {P}}}$ denotes Cauchy principal value of the integral. 
In formula  (\ref{I}) to bypass the poles of the denominator, we use the regularization procedure by adding small imaginary part $-i\varepsilon$, with $\varepsilon\rightarrow+0$. 
Momentum space wave function is related to the coordinate space wave function by the transformation
\be
\psi(x)=\int_{-b}^{b}e^{\frac{ipx}{\hbar}}\varphi(p)dp.
\ee
The integral 
\be
\int_{-b}^{b}\frac{e^{\frac{ipx}{\hbar}}}{g^2(p)-k^2}dp
\ee
have to be regularized in a similar way as in (\ref{I}) 
\be\label{int}
\int_{-b}^{b}\frac{e^{\frac{ipx}{\hbar}}}{g^2(p)-k^2-i\varepsilon}dp, 
\ee
with $\varepsilon\rightarrow+0$.
We obtain (see {\it{Appendix B}}) that in the limit $|x|\rightarrow\infty$ we can write
\be\label{assymp}
\int_{-b}^{b}\frac{e^{\frac{ipx}{\hbar}}}{g^2(p)-k^2-i\varepsilon}dp\rightarrow
 \frac{i\pi}{kf(k)}e^\frac{ip_0|x|}{\hbar}.
\ee
Thus, coordinate space wave function in the same limit  $|x|\rightarrow\infty$ has the form
\be
\psi(x)=e^\frac{ip_0x}{\hbar}+\frac{i\pi A}{kf(k)}e^\frac{ip_0|x|}{\hbar}
\ee
Using the expression for flow probability density (\ref{flux}) we obtain transition and reflection coefficients
\be
T=\frac{J_t}{J_0}=\left|1+\frac{i\pi A}{kf(k)}\right|^2=\frac{k^2f^2(k)\left(1+\tilde{V}_0G(k)\right)^2}{k^2f^2(k)\left(1+\tilde{V}_0G(k)\right)^2+{\pi^2V_0^2}},
\ee
\be
R=\frac{J_r}{J_0}=\left|\frac{i\pi A}{kf(k)}\right|^2=\frac{{\pi^2\tilde{V}_0^2}}{k^2f^2(k)\left(1+\tilde{V}_0G(k)\right)^2+{\pi^2\tilde{V}_0^2}}.
\ee
Note that $T+R=1$, as it have to be.

Let us consider some special examples of deformation function.

\textbf{Example 1.} 
In case of undeformed algebra 
\be f(P)=1,\ \  g(p)=p,\ \  b=\infty\ee
integral $G(k)$ is equal to zero
 \be G(k)={\displaystyle {\mathcal {P}}} \int_{-\infty}^\infty\frac{1}{p^2-k^2}dp=0.\ee
 We arrive to well-known undeformed  transition and reflection  coefficients
 \be T=\frac{k^2}{k^2+{\pi^2 \tilde{V}_0^2}}, \ \  R=\frac{\pi^2\tilde{V}_0^2}{k^2+{\pi^2 \tilde{V}_0^2}}.\ee
 
\textbf{Example 2.} 
 In the simplest case of deformation function, which corresponds to cutoff procedure in momentum space
 \be f(P)=1,\ \  g(p)=p,\ \  b<\infty\ee
  integral (\ref{pv}) is equal to
 \be G(k)={\displaystyle {\mathcal {P}}} \int_{-b}^b \frac{1}{p^2-k^2}dp=\frac{1}{k}\ln\left(\frac{b-k}{b+k}\right).\ee
 Corresponding transition and reflection  coefficients are
 \be\label{cutoff} T=\frac{k^2\left(1+\tilde{V}_0G(k)\right)^2}{k^2\left(1+\tilde{V}_0G(k)\right)^2+{\pi^2\tilde{V}_0^2}}, \ \  R=\frac{{\pi^2\tilde{V}_0^2}}{k^2\left(1+\tilde{V}_0G(k)\right)^2+{\pi^2\tilde{V}_0^2}}.\ee
 
 From the formula (\ref{cutoff}) for transition  coefficients we arrive at  a conclusion that  in case of the incident wave energy satisfying
 \be1+\tilde{V}_0G(k)=1+\frac{\tilde{V}_0}{k}\ln\left(\frac{b-k}{b+k}\right)=0,\ee
 we have perfect reflection. 
 
\begin{figure}[h]
 	\centering
 	\includegraphics[width=12 cm]{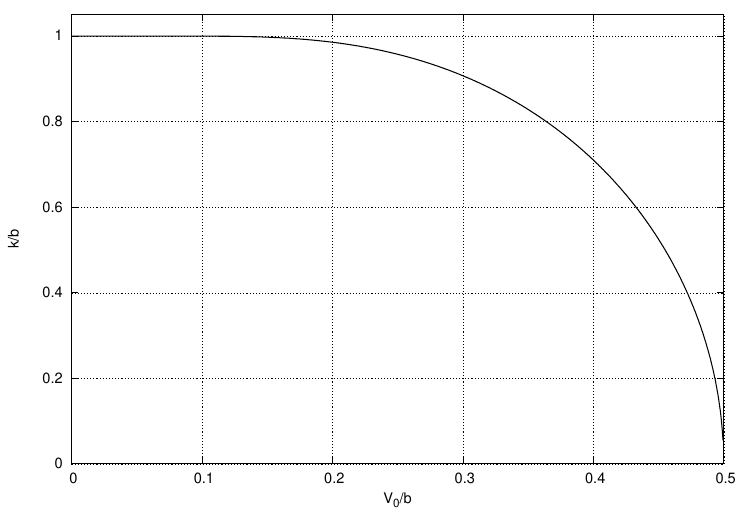}
 	\vspace{-10 pt}
 	\caption{\footnotesize{Reflection resonance energy on coupling constant for Example 2 of deformation function}}
 	\label{fig2}
 \end{figure}

 Dependence of reflection resonance energy $k$  on coupling constant $V_0$ is presented on Fig. \ref{fig2}.

\textbf{Example 3.} 
\begin{figure}[h]
	\centering
	\includegraphics[width=12 cm]{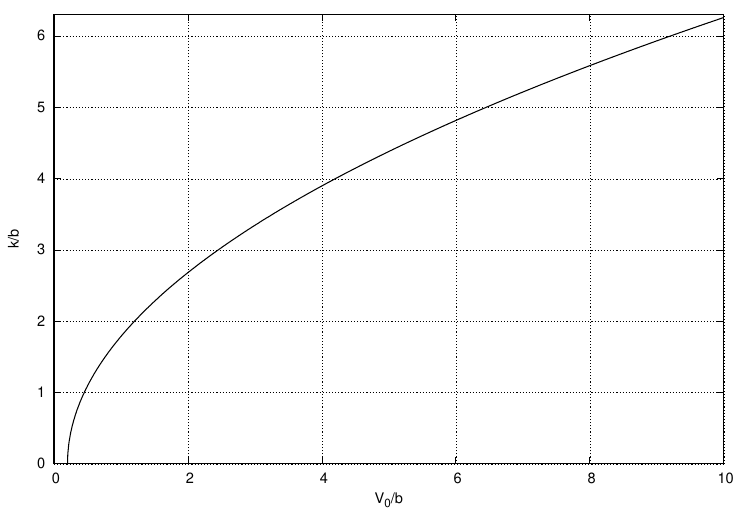}
	\vspace{-10 pt}
	\caption{\footnotesize{Reflection resonance energies depended on coupling constant for Example 3 of deformation function}}
	\label{fig3}
\end{figure}
In case  of deformation function proposed by Kempf 
 \be f(P)=1+\beta P^2, \ \ g(p)=\frac{1}{\sqrt{\beta}}\tan{\sqrt{\beta}p},\ \  b=\frac{\pi}{2\sqrt{\beta}}\ee
 we obtain
 \be G(k)={\displaystyle {\mathcal {P}}} \int_{-\frac{\pi}{2\sqrt{\beta}}}^\frac{\pi}{2\sqrt{\beta}} \frac{1}{\frac{1}{{\beta}}\tan^2{\sqrt{\beta}p}-k^2}dp=-\frac{\pi\sqrt{\beta}}{1+\beta k^2}.\ee
 Transition and reflection  coefficients are 
  \be\label{cutoff} T=\frac{k^2\left(1-\sqrt{\beta}\pi\tilde{V}_0+\beta k^2\right)^2}{k^2\left(1-\sqrt{\beta}\pi\tilde{V}_0+\beta k^2\right)^2+{\pi^2\tilde{V}_0^2}}, \ \  R=\frac{{\pi^2\tilde{V}_0^2}}{k^2\left(1-\sqrt{\beta}\pi\tilde{V}_0+\beta k^2\right)^2+{\pi^2\tilde{V}_0^2}}.\ee
Again, we have perfect reflection  in case of the incident wave energy satisfying
 \be1+\tilde{V}_0G(k)= 1-\tilde{V}_0\frac{\pi\sqrt{\beta} }{1+\beta k^2}=0.\ee
 
However, the  dependence of the reflection resonance energy $k$  on coupling constant $\tilde{V}_0$ (see Fig. \ref{fig3}) is quite different to previous example of deformation function.

 \textbf{ Example 4.} 
 Finally, we consider the case of deformation function with maximal momentum
  \be f(p)=\sqrt{1-\beta p^2}, \ a=\frac{1}{\sqrt{\beta}}; \ \ \  g(p)=\frac{1}{\sqrt{\beta}}\sin(\sqrt{\beta}p), \ \ b=\frac{\pi}{2\sqrt{\beta}}.\ee
  
    We obtain the following  for  integral (\ref{pv}) 
  \be  G(k)={\displaystyle {\mathcal {P}}}
   \int_{-\frac{\pi}{2\sqrt{\beta}}}^\frac{\pi}{2\sqrt{\beta}} \frac{1}{\frac{1}{{\beta}}\sin^2{\sqrt{\beta}p}-k^2}dp={\displaystyle {\mathcal {P}}} \int_{-\frac{1}{\sqrt{\beta}}}^\frac{1}{\sqrt{\beta}} \frac{1}{(P^2-k^2)\sqrt{1-\beta P^2}}dP=0,\ee 
  
  and transition and reflection  coefficients
  \be
  T=\frac{k^2(1-\beta k^2)}{k^2(1-\beta k^2)+{\pi^2\tilde{V}_0^2}}, \ \   R=\frac{{\pi^2\tilde{V}_0^2}}{k^2(1-\beta k^2)+{\pi^2\tilde{V}_0^2}}.
  \ee
  
  Perfect reflection occurs only for the maximally allowed energy of the incident wave $k=\frac{1}{\sqrt{\beta}}$. 
\section{Conclusion}
In present paper we have considered the  Dirac $\delta$-function potential problem
in general case of deformed Heisenberg algebra leading to the minimal length. 
The problem of definition of $\delta$-potential in deformed space has been examined. 
We have argued that usual definition of $\delta$-potential as usual Dirac $\delta$-function is inconsistent with the deformed space with minimal length.  In our previous paper \cite{Samar1}, we proposed the definition of the delta potential assuming that the kernel of the potential energy operator does not change in the case of deformation. In present paper we  have obtained that in quasiposition representation this definition  corresponds to more complicated one (\ref{delta}) containing operator of projection onto the origin. 

In quasiposition representation we have presented exact bound solutions of the problem. We have obtained the condition for energy spectrum, which similarly to undeformed case consists of one energy level.  Analysing the dependence of the eigenfunction in quasiposition representation on deformation parameter $b$ we have concluded that increasing this parameter  the eigenfunction  became more sharp at the origin reaching the features of undeformed eigenfunction.

We have adapted the continuity equation, which was exactly derived for the general case of deformed space with minimal length in \cite{Laba}, onto the case of potential energy operator containing the projection  operator. In the expression for the flux we have obtained additional term which can be considered as the neglectable one "outside" the point interaction.

We have solved exactly scattering problem for delta potential and have derived the expressions for reflection and transition coefficients. We have obtained  that for some resonance energy  the incident wave is completely reflected. We have concluded that this effect is very sensitive to the choice of  deformation function.

\section{Acknowledgement}
This work was supported by the Project by the Project 2020.02/0196 (No. 0120U104801) from National Research Foundation of Ukraine.

\section {Data availability}
The data that support the findings of this study are available from the corresponding author upon reasonable request.

\newpage
\appendix
\section{On the inconsistence of usual  $\delta$-potential  with the deformed space with minimal length}
\numberwithin{equation}{section}
\setcounter{equation}{0}
Let us consider the particle of mass $m$ in a delta well in space with minimum length.
In a quasi-coordinate representation, the potential energy will be considered to be the same as in the undeformed case
\be \label{naive_delta} V(x)=-V_0\delta(x). \ee
Let us write the Schr\"odinger equation for the considerable problem 
\be
\left(g^2{\left(-i\hbar\frac{d}{dx}\right)}+ q^2\right)\phi(x)-2m V_0\delta(x)\phi(x)=0.
\ee

Here we prove that the equation (\ref{incorrect_eqn}) has only trivial solution.
Without loss of generality we propose the solution  of equation (\ref{incorrect_eqn}) as
\be\label{incorrect_sol}
\phi(x)= \int_{-b}^{b}{\frac{C(p)e^{i px/\hbar}}{g^2(p)+q^2}dp}.
\ee
To belong to the domain of the operator of kinetic energy  wave function (\ref{incorrect_sol}) must be smooth.
Substituting (\ref{incorrect_sol}) into (\ref{incorrect_eqn}) we obtain
\be\label{incorrect_eqn2}
\int_{-b}^{b}{C(p)e^{i px/\hbar}}dp -2m V_0\phi(0)\delta(x)=0.
\ee
Here we use the property of $\delta$-function $\delta(x)\phi(x)=\delta(x)\phi(0)$.
Substituting in (\ref{incorrect_eqn2}) the following integral representation of $\delta$-function 
\be
\delta(x)=\frac{1}{2\pi\hbar}\int_{-\infty}^{\infty}dpe^{i px/\hbar}
\ee
we obtain
\be\label{incorrect_eqn3}
\int_{-b}^{b}{C(p)e^{i px/\hbar}}dp -\frac{m V_0\phi(0)}{\pi\hbar}\int_{-\infty}^{\infty}e^{i px/\hbar}dp=0.
\ee
Equation (\ref{incorrect_eqn3}) can be rewritten as the following 
\be\label{incorrect_eqn4}
\int_{-b}^{b}{\left(C(p)-\frac{m V_0\phi(0)}{\pi\hbar}\right)e^{i px/\hbar}}dp -\frac{m V_0\phi(0)}{\pi\hbar}\left(\int_{-\infty}^{-b}e^{i px/\hbar}dp+\int_{b}^{\infty}e^{i px/\hbar}dp\right)=0.
\ee
Since  complex exponentials $\frac{1}{\sqrt{2\pi\hbar}}e^{i px/\hbar}$  form an orthogonal basis, equation (\ref{incorrect_eqn3}) has only trivial solution $C(p)=0$.
This fact means that the problem (\ref{incorrect_eqn3}) can not be considered as the correctly defined one.

It is worth to notice, that in paper \cite{Gusson} the same problem was considered in the first order of deformation parameter $\beta$ 
\be\label{incorrect_approx}
-\frac{\hbar^2}{2m}\frac{d^2\phi(x)}{dx^2}+\beta\frac{\hbar^3}{3m}\frac{d^4\phi(x)}{dx^4}-V_0\delta(x)\phi(x)=E\phi(x).
\ee

The results obtained in \cite {Gusson} are generated by the usage of incorrect approximation because, as seen above, the problem has no exact solutions.

\section{Derivation of formula (\ref{assymp})}
\numberwithin{equation}{section}
\setcounter{equation}{0}
In case of $x>0$ we make the change of variables $u=px$  in (\ref{int})
\be
\int_{-b}^{b}\frac{e^{\frac{ipx}{\hbar}}}{g^2(p)-k^2-i\varepsilon}dp=\frac{1}{x}\int_{-R}^{R}\frac{e^{\frac{iu}{\hbar}}}{g^2(u/x)-k^2-i\varepsilon}du, 
\ee
with $R=bx$.

\begin{figure}[h]
	\centering
	\includegraphics[width=10 cm]{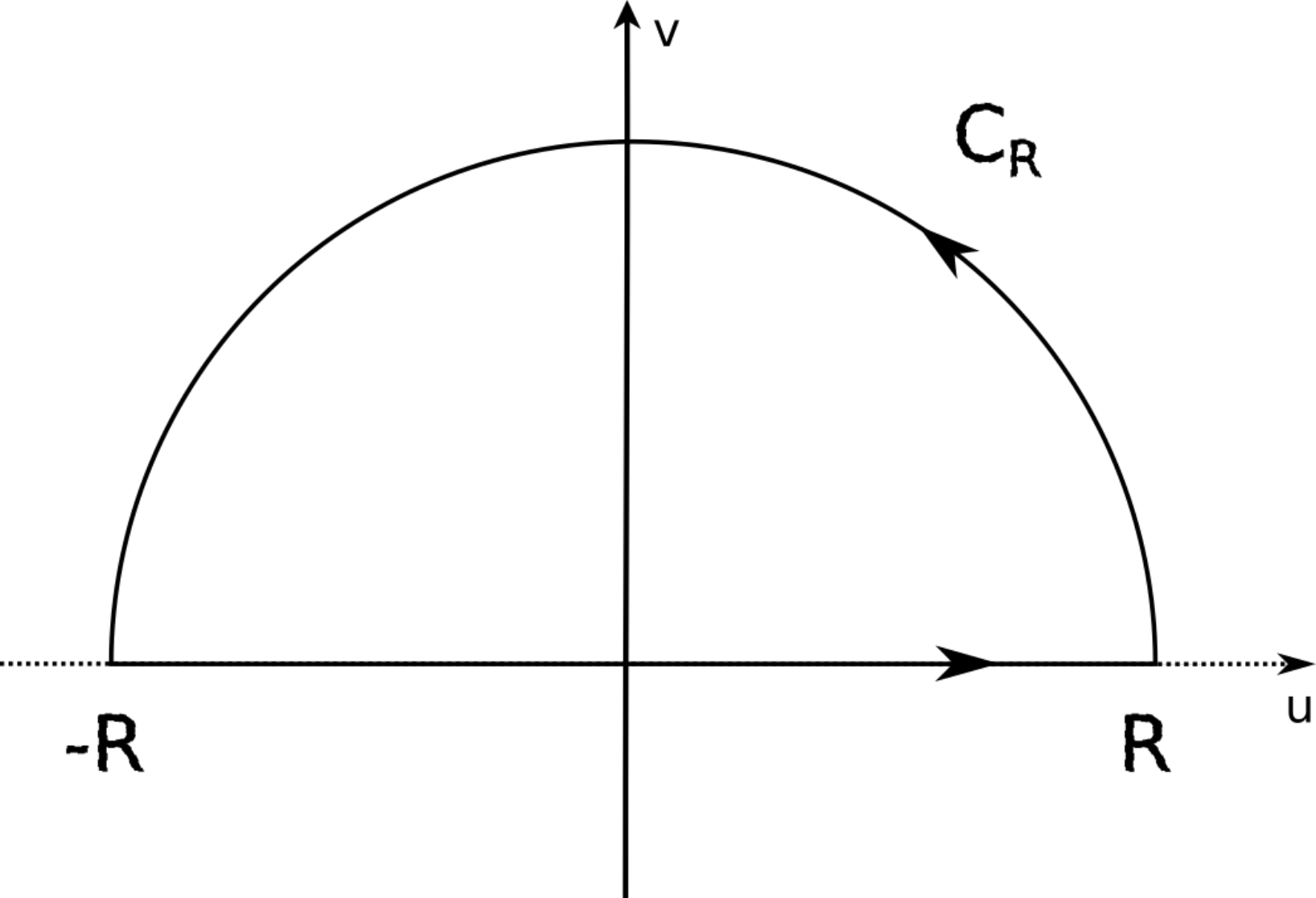}
	\vspace{-10 pt}
	\caption{\footnotesize{Reflection resonance energies depended on coupling constant}}
	\label{fig4}

\end{figure}

According to the Cauchy's residue theorem
\be\label{Cauchy}
\frac{1}{x}\int_{-R}^{R}\frac{e^{\frac{iu}{\hbar}}}{g^2(u/x)-k^2-i\varepsilon}du+\frac{1}{x}\int_{C_{R}}\frac{e^{\frac{iz}{\hbar}}}{g^2(z/x)-k^2-i\varepsilon}dz=2\pi i\sigma,
\ee
where $z=u+iv$ and $\sigma$ being the sum of residues of the function under integration at the singular points inside the contour of integration given on Fig.\ref{fig4}.

In the limit $x\rightarrow +\infty$ the second integral in (\ref{Cauchy}) goes to zero due to Jordan's lemma.
The residue in the pole can be easily obtained
\be\sigma=\frac{1}{2kg'(k)}e^\frac{ip_0x}{\hbar}= \frac{1}{2kf(k)}e^\frac{ip_0x}{\hbar},\ee
with $p_0$ is given by $g(p_0)=k$.
Finally, in the limit $x\rightarrow +\infty$ we obtain
\be
\int_{-b}^{b}\frac{e^{\frac{ipx}{\hbar}}}{g^2(p)-k^2-i\varepsilon}dp= \frac{i\pi}{kf(k)}e^\frac{ip_0x}{\hbar}.
\ee
Similarly, in case of $x<0$ in the limit of $x\rightarrow -\infty$
\be
\int_{-b}^{b}\frac{e^{\frac{ipx}{\hbar}}}{g^2(p)-k^2-i\varepsilon}dp= \frac{i\pi}{kf(k)}e^\frac{-ip_0x}{\hbar}.
\ee
Combining these two results, we obtain formula (\ref{assymp}).


\newpage
\begin{thebibliography}{999}
	
	\bibitem{Maslowski} T. Maslowski, A. Nowicki and V. M. Tkachuk, J. Phys. A 45, 075309 (2012).
	\bibitem{Kempf1994}A. Kempf, J. Math. Phys. 35, 4483 (1994).
	\bibitem{KempfManganoMann} A. Kempf, G. Mangano  and R. B. Mann, Phys. Rev. D 52, 1108 (1995).
	\bibitem{HinrichsenKempf}H. Hinrichsen  and A. Kempf, J. Math. Phys. 37, 2121 (1996).
	\bibitem{Kempf1997} A. Kempf, J. Phys. A 30, 2093 (1997).
	\bibitem{GrossMende}D. J. Gross  and P. F. Mende,  Nucl. Phys. B 303, 407 (1988).
	\bibitem{Maggiore}M. Maggiore,  Phys. Lett. B 304, 65 (1993).
	\bibitem{Witten}E. Witten,  Phys. Today 49, 24 (1996).
	\bibitem{Brau} F. Brau, J. Phys. A 32, 7691 (1999).
	\bibitem{Benczik}S. Benczik, L. N. Chang, D. Minic and T. Takeuchi, Phys. Rev. A 72, 012104 (2005).
	\bibitem{StetskoTkachuk}M. M. Stetsko and V. M. Tkachuk, Phys. Rev. A 74, 012101 (2006).
	\bibitem{Stetsko2006} M. M. Stetsko, Phys. Rev. A 74, 062105 (2006).
	\bibitem{Stetsko2008}M. M. Stetsko and V. M. Tkachuk, Phys. Lett. A 372, 5126 (2008).
	\bibitem{SamarTkachuk}M. I. Samar and V. M Tkachuk, J. Phys. Stud. 14, 1001 (2010).
	\bibitem{Samar} M. I. Samar, J. Phys. Stud. 15, 1007 (2011).
	
	\bibitem{Brau2006}F. Brau, F. Buisseret, Phys. Rev. D 74, 036002, (2006).
	\bibitem{Nozari2010} K. Nozari, P. Pedram, Europhys. Lett. 92, 50013 (2010).
	\bibitem{Pedram2011} P. Pedram,  K.Nozari, S.H.Taheri, J. H. En. Phys. 1103, 093 (2011)
	
	
	
	
	\bibitem{Ferkous} N. Ferkous, Phys. Rev. A 88, 064101 (2013).
	\bibitem{Samar1} M. I. Samar, V. M Tkachuk, J. Math. Phys. 57, 042102 (2016).
	\bibitem{Samar2} M. I. Samar, V. M Tkachuk, J. Math. Phys. 57, 082108 (2016).
	\bibitem{Fityo} T. V. Fityo, I. O. Vakarchuk and V. M. Tkachuk, J. Phys. A 39, 2143 (2006).
	
	
	\bibitem{Bouaziz2007}D. Bouaziz and M. Bawin, Phys. Rev. A 76, 032112 (2007).
	\bibitem{Bouaziz2008} D. Bouaziz and M. Bawin, Phys. Rev. A 78, 032110 (2008).
	\bibitem{Bouaziz2017}D. Bouaziz and T. Birkandan, Ann.  Phys. 387, 62 (2017).
	\bibitem{Samar2020} M. I. Samar, V. M Tkachuk, J. Math. Phys. 61, 092101 (2020).
	
	\bibitem{Gusson} M. F. Gusson, A. O. O. Gonçalves, R. O. Francisco  et al., Eur. Phys. J. C 78, 179 (2018).

	
	\bibitem{Laba} H. P. Laba, V. M. Tkachuk, Phys. Lett. A 391, 127141 (2021).
	
	
\end{thebibliography}
\end{document}